# Assessment of the (010) β-Ga$_2$O$_3$ Surface and Substrate Specification

M.A. Mastro [1, a)], C.R. Eddy Jr., [1], M.J. Tadjer [1], J.K. Hite [1], J. Kim [2], and S.J. Pearton [3]

[1] U.S. Naval Research Laboratory, Washington, DC 20375, USA
[2] Department of Chemical and Biological Engineering, Korea University, Seoul 02841, Korea
[3] Department of Materials Science and Engineering, University of Florida, Gainesville, FL 32611, USA

**Abstract**

Recent breakthroughs in bulk crystal growth of the thermodynamically stable beta phase of gallium oxide (β-Ga$_2$O$_3$) have led to the commercialization of large-area β-Ga$_2$O$_3$ substrates with subsequent epitaxy on (010) substrates producing high-quality films. Still, metal-organic chemical vapor deposition (MOCVD), molecular beam epitaxy (MBE), and processing of the (010) β-Ga$_2$O$_3$ surface are known to form sub-nanometer scale facets along the [001] direction as well as larger ridges with features perpendicular to the [001] direction. A density function theory calculation of the (010) surface shows an ordering of the surface as a sub-nanometer-scale feature along the [001] direction. Additionally, the general crystal structure of β-Ga$_2$O$_3$ is presented and recommendations are presented for standardizing (010) substrates to account for and control the larger-scale ridge formation.

**INTRODUCTION**

β-Ga$_2$O$_3$ possesses fundamental electronic properties that make transistors and diodes fabricated on this material well-suited for high power devices. A number of these properties directly derive from the wide band-gap of β-Ga$_2$O$_3$ ($E_g$ = 4.85 eV) including an exceptionally high electric breakdown field (approximately 8 MV/cm).[1-4] This high breakdown field allows β-Ga$_2$O$_3$-based devices to be biased at a high drain voltage while maintaining a large dynamic range.[5-7] Furthermore, the wide band-gap of β-Ga$_2$O$_3$ allows device operation at elevated temperature.[8-10] Amongst all wide-bandgap semiconductors, including GaN and SiC, a key advantage for Ga$_2$O$_3$ is the current and projected low-cost of β-Ga$_2$O$_3$ substrates.[11, 12]

Examining the structure of the [010] oriented β-Ga$_2$O$_3$ crystal helps to frame a number of issues in the epitaxy on and behavior of (010) Ga$_2$O$_3$. The crystal structure of β-Ga$_2$O$_3$ is a base-centered monoclinic (b-axis unique, space group 12, a = 12.214 Å, b = 3.0371 Å, c = 5.8981 Å). Inspecting the structure of the β-Ga$_2$O$_3$ crystal in Fig. 1 shows that the Ga$^{3+}$ cations have two distinct bonding coordination. The Ga (I) cation forms a slightly distorted tetrahedral coordination with four O ions and the Ga (II) cations has a highly distorted octahedral coordination with six O ions. Three inequivalent O sites are present. The O(I) is threefold coordinated to two octahedral coordinated Ga ions and one tetragonal Ga ion; the O(II) is threefold coordinated to one octahedral coordinated Ga ion and two tetragonal Ga ions; the O(III) is fourfold coordinated to three octahedral coordinated Ga ions and one tetragonal Ga ion.[13]

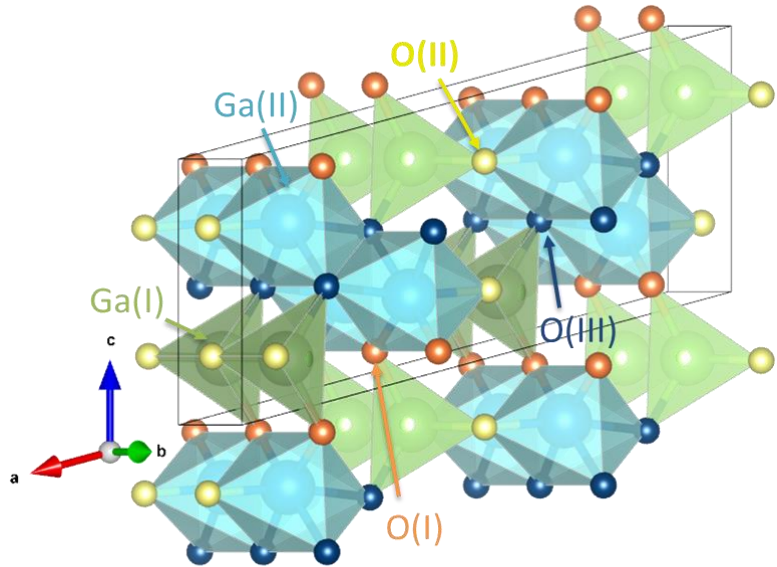

FIG. 1. Bonding structure of monoclinic (C2/M group symmetry) β-$Ga_2O_3$ phase. The Ga (I) cation (green in online version) bonds to four O ions and the Ga (II) cations (blue in online version) bonds to six O ions. Each O(I) (orange in online version) bonds to two octahedral coordinated Ga ions and one tetragonal Ga ion. Each O(II) (yellow in online version) bonds to one octahedral coordinated Ga ion and two tetragonal Ga ions. Each O(III) (dark blue in online version) bonds to three octahedral coordinated Ga ions and one tetragonal Ga ion.

The crystal is oriented (roughly in) Fig. 1 and (exactly in) Fig. 2 such that the [010] b-axis projected towards the reader. The nature of the C2/M group symmetry that describes the β-$Ga_2O_3$ crystal sets the [010] b-axis as perpendicular to the (010) plane; in contrast, the [100] a-axis is not aligned with the perpendicular to the (100) plane, and the [001] c-axis is not aligned with the perpendicular to the (001) plane. Specifically, the a- and c- axis form an angle, β, of 103.68°.

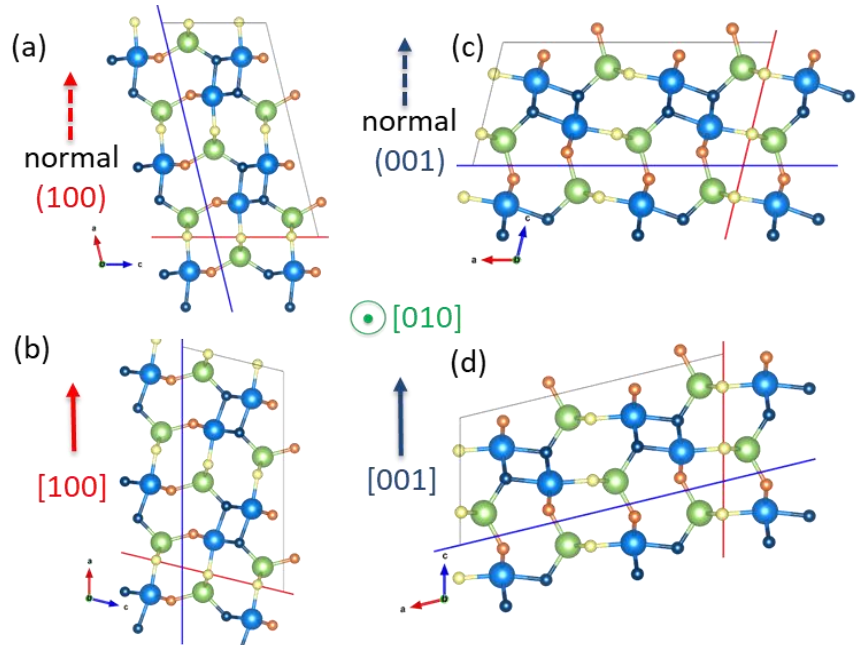

FIG. 2. Depiction of the (010) plane (parallel to the page), which highlights that the a- and c-directions are not perpendicular (β=103.68°). It is evident that, in (a), the perpendicular to the (100) a-plane and, in (b), the [100] direction are not aligned. Similarly, it is evident that, in (c), the perpendicular to the (001) a-plane and, in (d), the [001] direction are not aligned.

Remarkable progress has been made over the past few years in the bulk crystal growth of β-$Ga_2O_3$.[13] Edge-defined film-fed growth (EFG) has led to the production of up to 4-inch diameter $Ga_2O_3$ substrates. These results were attained by pulling the boule along the [010] direction at a rate of approximately 15mm/hr. Similarly, bulk Czochralski growth of β-$Ga_2O_3$ is commonly conducted along the [010] direction wherein any deviation from the [010] direction will result in cracks along the primary (100) and secondary (001) cleavage planes. In Czochralski bulk growth, the weight of the crystal and temperature gradient of the process exacerbates cleavage along these planes.[14]

The (010) β-$Ga_2O_3$ substrate is a common choice for epitaxy as these substrates are commercially available and, as shown by Sasaki et al., yield growth rates much higher for MBE relative to other planes (particularly compared to the (100) cleavage plane).[15] Moreover, recent studies of MOCVD growth on (010) β-$Ga_2O_3$ substrates have shown that this orientation is amenable to producing low impurity films with a high degree of structural order including a recent MOSFET with an n⁺ doped channel that displayed a mobility of 170 $cm^2$/V·s at room temperature.[16]

As discussed by Mazzolini and Bierwagen,[17] closer examination of the (010) β-$Ga_2O_3$ surface reveals two type of features on the (010) surface. First, thermodynamically determined sub-nanometer facets defined by the $(\bar{1}10)$ and (110) plane form as elongated pyramidal stripes in the [001] direction with a lateral (i.e., perpendicular to [001]) spacing of 5 to 10 nm. Mazzolini et al. provided the initial analysis on the stabilization of these facets under metal-rich growth conditions.[18] Second, trenches are observed with features orthogonal to the [001] direction with a trench to trench spacing 300 to 500 nm and a depth of 5 to 10 nm. Mazzolini and Bierwagen showed that an (unintentional) offcut of approximately 0.1° towards the [001] direction enables step-flow growth, which prevents the formation of the ridges.[17] A similar facet and ridge morphology is observed in (010) β-$Ga_2O_3$ thin-films deposited by metal-organic chemical vapor deposition (MOCVD) and hydride vapor phase epitaxy.[19-21]

This article discusses the general crystal of β-$Ga_2O_3$, presents DFT calculations of the β-$Ga_2O_3$

crystal with a focus on understanding the surface structure of the (010) surface, and provides a recommendation for standardizing substrate flats to control ridge formation during deposition of (010) β-Ga$_2$O$_3$.

**MODELLING**

To provide insight into the bonding structure, density of states, and band structure of the β-Ga$_2$O$_3$, a density functional theory (DFT) calculation was conducted via Quantum Espresso.[22, 23] The Perdew-Burke-Ernzerhof (PBE) scheme described the exchange-correlation energy where the ultrasoft pseudo-potential method described the electron–ion interactions. The General Gradient Approximation plus Hubbard U (GGA + U) formalism set the Hubbard U parameters as 7.0 eV and 8.5 eV for Ga and O ions, respectively, to accurately calculate the bandgap. The Broyden-Fletcher-Goldfarb-Shanno (BFGS) optimization method with an energy cutoff of 450 eV was used to calculate the structural relaxation.

**RESULTS**

*Analysis (010) Surface*

Figure 3 shows the (010) slab in the bulk and relaxed surface state with cross-sections displayed for both the [001] and [100] direction coming out of the page. The first two monolayers are mobile while the remaining atoms are fixed as an approximation to a bulk structure. Close examination of the relaxed crystal in Fig. 3(b) reveals significant atom movement perpendicular to the [001] direction; in contrast little atom movement is visible in Fig. 3(d). That is, the surface relaxes such that monolayer features form with a long-axis along the [001] direction.

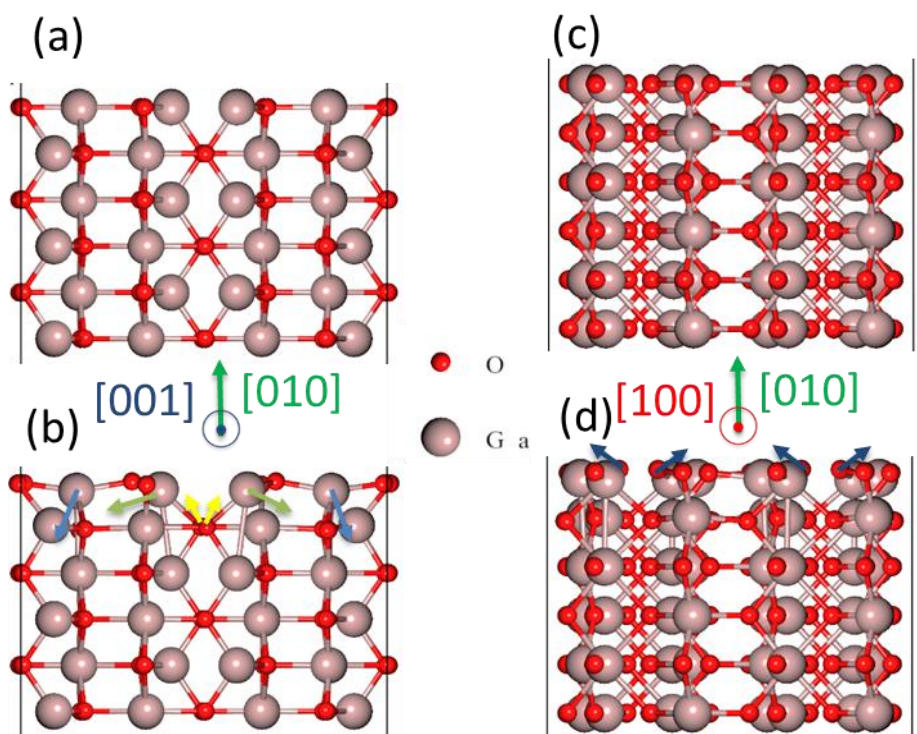

FIG. 3. Crystal structure of (010) β-Ga$_2$O$_3$ slab (with a coloration of light-brown Ga ions and red O ions, following the Jmol convention, in online version) for comparison to Fig. 4. Cross-sections are given for the (a) bulk crystal and (b) crystal with the calculated relaxed top surface with the [001] direction out of the screen. Similarly, the cross-section for the (c) bulk crystal and (d) crystal with the calculated relaxed top surface for [100] direction out of the screen. The direction (i.e., not magnitude) of atomic motion in relaxation are displayed, for visual clarity, only for a limited set of atoms. For assignment of the atomic coordination see alignment with Figs. 5(b) and 5(d)

A reconstruction of the surface will naturally create states distinct from the bulk density of states. In Fig. 4(b), the projected states from the (light-brown in online version) Ga orbitals and (red in online version) O orbitals are shown for bulk β-Ga$_2$O$_3$ and the surface of a relaxed (010) β-Ga$_2$O$_3$ slab. Specifically, the local projected states from the first two, relaxed monolayers are shown. Closer examination of the states at the lower range of density (i.e., the y-axis) show a presence of states, albeit small, within the gap near the conduction band edge.

For reference, Fig. 4(a) contains the density of states aligned to the band structure of bulk β-Ga$_2$O$_3$. The valence band maximum in β-Ga$_2$O$_3$ forms from weakly interacting O 2p orbital states with contribution of Ga 3d and 4s orbitals.[24] The reader is directed to Peelaers and C. G. Van de Walle[25] for a detailed description of the Brillouin zone and band structure of the β-Ga$_2$O$_3$.

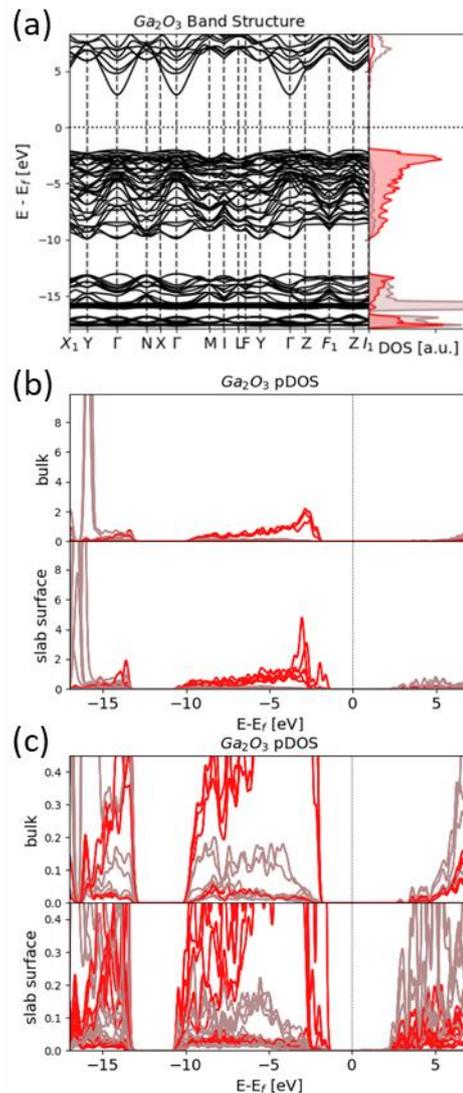

FIG. 4. (a) Bulk β-Ga$_2$O$_3$ band structure and corresponding density of states projected from the (light-brown in online version) Ga orbitals and (red in online version) O orbitals. (b) Comparison of projected density of states for bulk β-Ga$_2$O$_3$ to the local states from surface atoms in a relaxed state for (010) β-Ga$_2$O$_3$ slab. (c) Same projected density of state comparison for at the lower range (on the y-axis) shows new states within the bandgap near the conduction band edge.

*Ridge Formation*
Two types of surface features with a major-dimension along the [001] direction, as discussed above, were seen for MBE of (010) β-Ga$_2$O$_3$ films as well as after a surface clean or etch of the (010) β-Ga$_2$O$_3$ surface.[17, 18] Baldini et al. also found that MOCVD produces a (010) β-Ga$_2$O$_3$ surface with a morphology dominated by elongated islands oriented along the [001] direction.[21] Figure 5(a) depicts the crystal structure of a cross-section of a [001] ridge. The surface termination of this cross-section of the [001] ridge follows the experimental assignment of (Fig. 8 in) Mazzolini et al.[18] The corresponding bulk slab in Fig. 5(b) shows that these $(\bar{1}10)$ and (110) planes form a 13.9056° angle with the (010) plane.

The literature contains other references to surface features on the (010) surface. Rafique et al. labeled features as forming along the [100] direction.[26] Feng et al. stated that groove structures were along the (001) crystal orientation.[27] Confirmation of these planes is provided by extracting the facet profile from an AFM scan or similar technique. The lowest index inclined plane for a [100] ridge are the $(01\bar{1})$ and (011) planes,[28] which is depicted in Figs. 5(c) and 5(d). The atomic coordination assignment in Fig. 5 follows the convention described in Bermudez.[29]

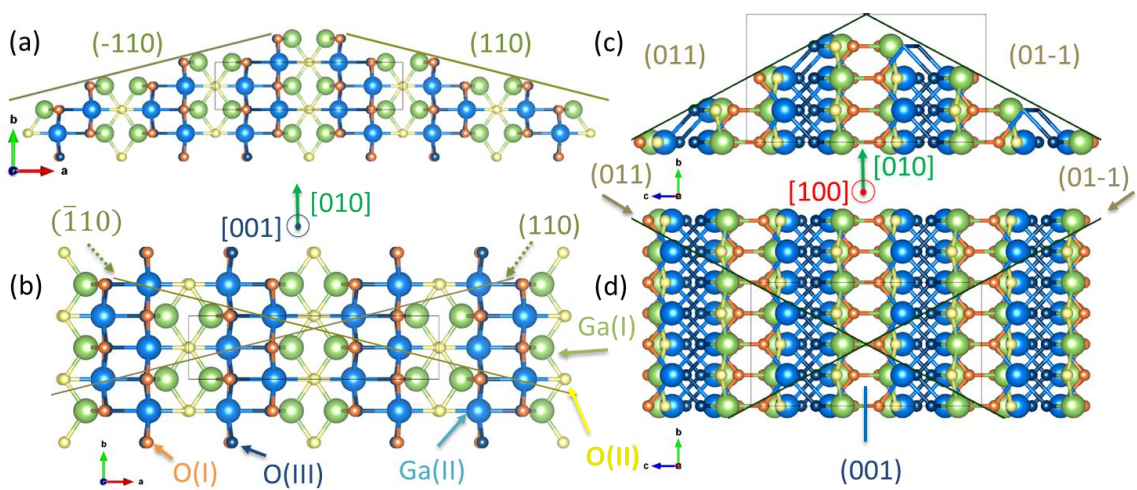

FIG. 5. Theoretical (unrelaxed) crystal structure of β-Ga$_2$O$_3$ [010] direction oriented upwards, i.e., the (010) plane is the top of each crystal ridge and corresponding slab. The two fundamental crystallographic orientation directions are shown for comparison. In (a, b) with [001] direction out of the screen, the $(\bar{1}10)$ and (110) planes form an angle of 13.9056° with the (010) plane. In (c, d) with [100] direction out of the screen, the $(01\bar{1})$ and (011) planes form an angle of 27.6841° with the (010) plane.

It is useful to consider the reconstruction of the near-surface and surface atoms on a structure of Fig. 5(a) composed of the $(\bar{1}10)$ and (110) planes. Figure 6 displays a ridge along the [001] direction on a (010) slab - as a cross-section with the [001] direction coming out of the page. Specifically, Fig. 6(a) shows the unrelaxed crystal [i.e., similar to the theoretical crystal in Fig. 5(a)] and Fig. 6(b) shows the same crystal after the relaxation calculation. Examination of the relaxed crystal in Fig. 6(b) reveals significant atom movement perpendicular to the [001] direction. The surface relaxes such that monolayer features form with a long-axis along the [001] direction, which is similar to the (010) surface relaxation in Fig. 3(b).

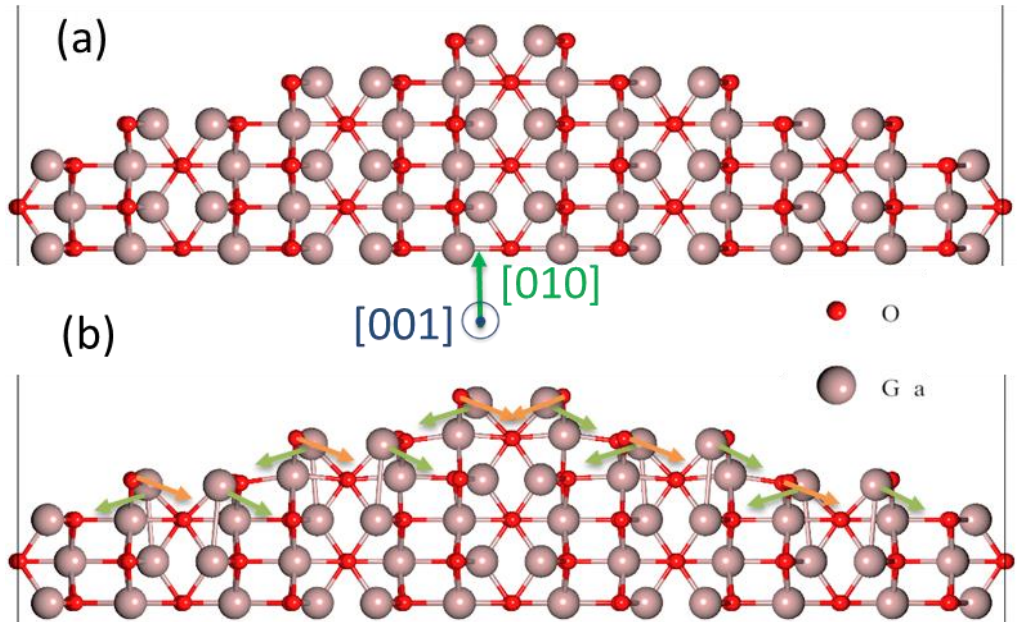

FIG. 6. Crystal structure of ridge oriented along the [001] direction on a (010) β-Ga$_2$O$_3$ slab (with light-brown Ga ions and red O ions in online version) are shown as cross-sections with the [001] direction out of the screen. (a) The unrelaxed crystal is similar to Fig. 5(a). (b) The relaxed crystal displays ordering of the ridge surface as sub-nanometer-scale features along the [001] direction, which is similar to the relaxation of a planar (010) surface that is depicted in Fig. 1(b). The direction (i.e., not magnitude) of atomic motion in relaxation are displayed, for visual clarity, only for a limited set of atoms. For assignment of the atomic coordination, see alignment with Fig. 5(a).

## DISCUSSION
### DFT Calculation

Above, a density function theory calculation of the (010) surface shows an ordering of the surface as a sub-nanometer-scale feature along the [001] direction. An important area of future work is studying the influence of chemical potential on the (010) surface. Mazzolini at al. showed that the deposition and annealing environment determines the atomic structure of the (010) surface.[18] More generally, it is important to understand how reducing or oxidizing environments affects the arrangement of the atomic structure (010) surface. Furthermore, it will be beneficial to connect the predicted atomic arrangement including resultant experimental findings including RHEED patterns.

### Proposed Substrate Specification
### Offcut Influence

This section provides further information on the influence of substrate offcut for deposition on the (010) β-Ga$_2$O$_3$ substrate and the Sec. IV B 2 provides a recommendation for standardization of the (010) β-Ga$_2$O$_3$ substrate. Mazzolini and Bierwagen stated that the trench formation arises from the asymmetry of the β-Ga$_2$O$_3$ (010) crystal surface, which creates a much longer diffusion length and lower nucleation density along the [001] direction.[17] Low or no offcut results in island growth that coalesces as trenches. An offcut along the [001] direction creates steps that act as periodic nucleation sites for step-flow growth – when the terrace width is shorter than the diffusion length of the Ga species. Okumura et al. demonstrated smooth (010) films by MBE where step-flow growth was attributed to the 2° offcut along the [001] direction.[30] For step-flow growth, the surface roughness will be determined by the sub-nanometer faceting of the $(\bar{1}10)$ and (110) planes.[17] This need to offcut relative to the [001] direction motivates an understanding of the optimal substrate geometry including wafer flat specifications for (010) β-Ga$_2$O$_3$ substrates.

*Substrate Standardization*

The EFG process has led to the production of up to 4-inch diameter $Ga_2O_3$ substrates oriented in $(\bar{2}01)$ or (001) planes. The float zone technique is able to produce (010) oriented boules with a diameter of 1-inch at a growth rate approaching 5mm/hr. The Czochralski technique has produced, at a pull rate of 2 mm/hr, boules of 2-inch diameter with potential for larger diameter boules in the future.[13] The dislocation density of current bulk wafers is of the order $10^3$ cm$^{-2}$, a key result for making large area power devices.[8, 11]

The EFG bulk growth process involves a high-purity $Ga_2O_3$ powder, which is inductively melted in an Ir crucible. A seed crystal in contact with the molten $Ga_2O_3$ is pulled through a slit to produce bulk plates with a top surface along $(\bar{2}01)$ or (001) planes. For the (010) face, the minimum attainable dimension, is defined by the height of the slit. The EFG process is less limited to the width of the slit, which is composed of the $(\bar{2}01)$ or (001) plane. The second dimension of the $(\bar{2}01)$ or (001) plane is only limited by the length (and thus growth time) of the plate.

The crystal with the $(\bar{2}01)$ top surface forms with the (100) plane. Rotation of the seed about the [010] growth direction allows the production of slabs with faces corresponding to the (100) and (001) planes. Common substrate structures produced via EFG are displayed in Fig. 7.

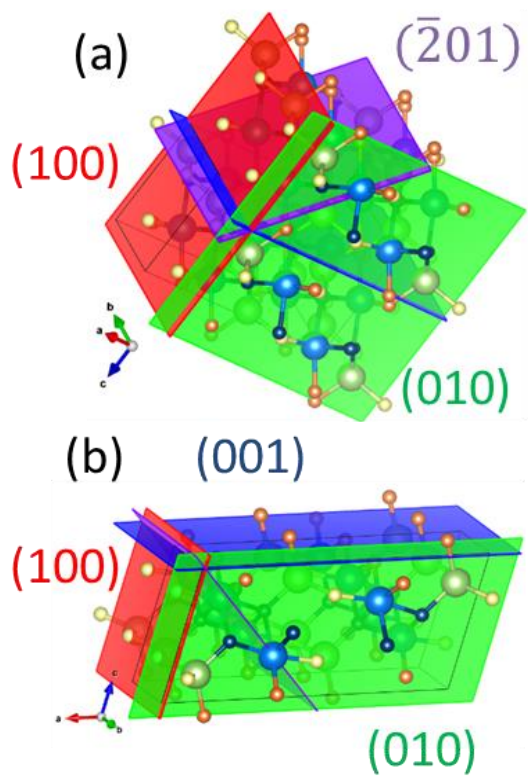

FIG. 7. Typical structure of β-$Ga_2O_3$ slabs produced by bulk EFG. The <010> pull direction results in production of slabs with a (010) (green) surface, which is approximately parallel to the screen. Depending on the rotation of the seed about the <010> pull direction, the second and third surface are either (a) $(\bar{2}01)$ (purple in online version) and (100) (red in online version), or (100) (red in online version) and (001) (blue in online version) planes. Orientation of the depictions in (a) and (b) is matched to experimental demonstration of ingots visible in Kuramata et al. (Refs 31 and 32, respectively).

The two current standard (010) β-$Ga_2O_3$ rectangular substrate geometries are shown in Fig. 8. The specification in Fig. 8(a) has one wafer edge in [102] direction and the other wafer edge as the perpendicular to [102] direction, which corresponds to the $(\bar{2}01)$ plane. This specification is designed to match the boule depicted in Fig. 7(a).

The second specification in Fig. 8(b) has one wafer edge set to the [001] direction and the second edge perpendicular to [001] direction. These directions are parallel and perpendicular, respectively, to the (100) plane. Thus, this specification was designed to match the boule shape depicted in Fig. 7(b). As discussed in Kuramata et al.,[32] the growth rate perpendicular to the (100) plane is low hence the facets are spontaneously formed at the (100) plane.

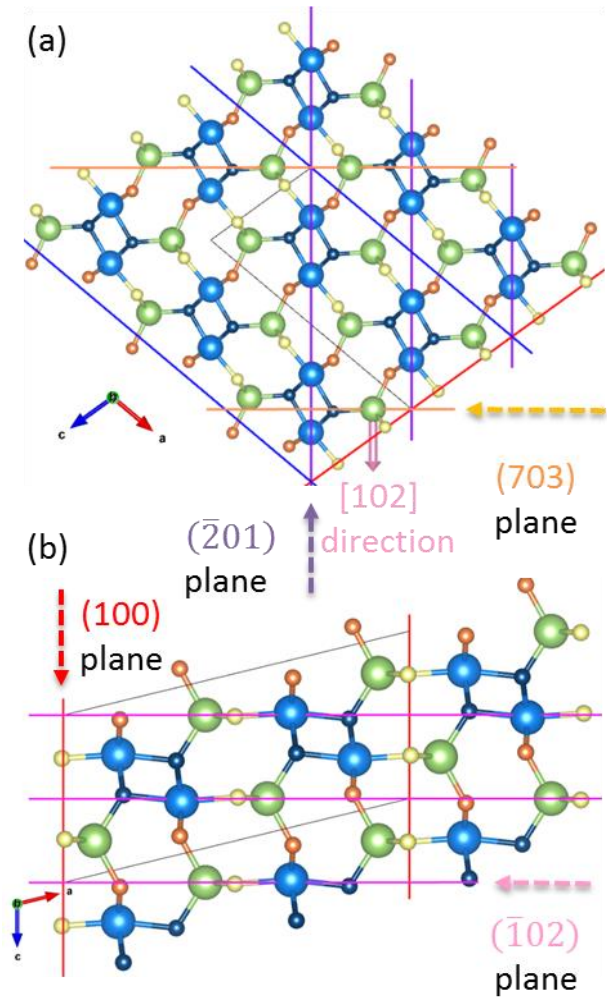

FIG. 8. Standard geometry of (010) β-$Ga_2O_3$ substrate. (a) The substrate specification for the substrate edges (and offset) is aligned to [102] direction and perpendicular to [102] direction. The [102] direction (oriented down in (a)) aligns to parallel to the $(\bar{2}01)$ plane and is perpendicular to the (703) plane. (b) The substrate specification for substrate edges (and offset) is aligned to the [001] direction and perpendicular to [001] direction. The [001] direction (oriented down in (b)) aligns parallel to (100) plane and perpendicular to the $(\bar{1}02)$ plane.

Given the importance as discussed above in defining offcut along the [001], the standard substrate specifications including wafer flats or edges should be structured to define this offcut. The (100) plane propensity to cleave is used to define sidewall of the bulk growth (as discussed above). Hence it is logical to use the (100) edge as the major flat or edge. The simplest approach is to use geometry of Fig. 8(b) composed of one edge (or major flat) as (100) (i.e., ∥ [001] direction) and the other edge (or minor flat) perpendicular to the [001] direction, i.e., the minor flat is rotated 90° from the major flat. This design sets the [001] offcut orthogonal to the (100) plane.

For comparison, the standard (0001) SiC substrate has a primary flat as the $(1\bar{1}00)$ plane, which is typically determined by a XRD back reflection technique. The edge of the primary flat is parallel to the $[11\bar{2}0]$ direction, and the minor flat, which is rotated 90° from the major flat, is the $(11\bar{2}0)$ plane. The standard offcut, e.g., 4°, is in the $[11\bar{2}0]$ direction, toward the minor flat.

It is important to revisit the question of the structure of the miscut substrate surface. A sawing process will leave a damaged surface and for offcut toward [001] with steps edges of or similar to $(\bar{1}02)$ (see Fig. 8(b)). A proper pre-growth treatment will force the damaged surface to reconstruct or decompose to create pristine step edges composed of (001) planes. Nevertheless, the (001) step-edges are not perpendicular to the offcut direction owing to the monoclinic structure. Examination of the (010) surface after growth in[16, 18, 31] and after an anneal in a Ga-rich environment (as in Mazzolini and Bierwagen[17]) reveals similar jagged morphology in the [001] direction. In other words, the surface in these varied studies do not form long-range step-edges formed from the (001) or similar plane.

Following this logic supports setting the offcut along the (001) normal (see Fig. 2(c)) – rather than along to the [001] direction (see Fig. 2(d)). This geometry should create (001) step edges – with a length theoretically up to length of the substrate. Specifically, the recommended substrate specification has an (100) major flat and a (001) minor flat separated by an angle, β, of 103.68°, and offcut is defined towards the (001) normal. This design is amenable to manufacture as it matches a standard wafering process where alignment to a plane is provided by x-ray diffraction.

## SUMMARY AND CONCLUSIONS

The wide bandgap and related electronic properties of β-$Ga_2O_3$ address a high-voltage application space not currently accessible with current semiconductor technology.[33-35] Combined with the projected availability of low-cost high-quality substrates, this material will additionally enable a new class of power devices for low-cost applications.[10] It is discussed that MOCVD and MBE on (010) β-$Ga_2O_3$ substrates produces a surface morphology defined by two types of features. The first feature is facets, oriented along the [001] direction, that are less than 10 nm in lateral extent with a vertical extent within 0.6 nm. A density function theory calculation shows (010) planar surface reconstructs with similar sub-nanometer features along the [001] direction. The second feature is ridges of greater than 10 nm in lateral extent and a vertical extent within (approximately) 8 nm. Prior MBE based studies showed that an (unintentional) small offcut along the [001] direction is necessary to create a periodic array of step edges to enable step-flow growth and thus eliminate the ridge morphology.[17] An analysis of the geometry of current (010) β-$Ga_2O_3$ substrate specifications is presented. The recommendation is a standard substrate specification with wafer edge (i.e., major flat) consisting of the (100) plane. One possible (and currently used for rectangular substrates) specification of a second edge in the [001] direction can be similarly specified for a 2-inch diameter wafer as a [001] minor flat. The future development of epitaxy on the (010) plane will be accelerated if (010) substrates are available with the (to be determined) optimal offcut. An argument is made that the (010) surface is more amenable to step-flow growth if this offcut is based on the normal to the (001) plane – as opposed to the [001] direction. In this design, the minor flat is the (001) plane separated by an angle, β, of 103.68° from the (100) major flat.

## ACKNOWLEDGMENTS

The work at NRL was partially supported by DTRA Grant No. HDTRA1-17-1-0011 (Jacob Calkins, monitor) and the Office of Naval Research. The work at UF is partially supported by HDTRA1-17-1-0011. The project or effort depicted is partially sponsored by the Department of the Defense, Defense Threat Reduction Agency. The content of the information does not necessarily reflect the position or the policy of the federal government, and no official endorsement should be inferred. The work at Korea University was supported by the Korea Institute of Energy Technology Evaluation and Planning (20172010104830) and the National Research Foundation of Korea (2020M3H4A3081799). The authors thank the anonymous reviewers for the beneficial information and analysis.

## DATA AVAILABILITY

The data that support the findings of this study are available from the corresponding author upon reasonable request.